\newcommand{\newapj}{} 
\newcommand{\includearxivid}{} 

\ifdefined\newapj
\documentclass[twocolumn]{aastex61}
	\ifdefined\includearxivid
	\else
	\fi
\else
\documentclass[preprint2]{emulateapj}
\fi

\usepackage{color}
\usepackage{amsmath}
\usepackage{threeparttable}
\usepackage{natbib}
\usepackage{graphicx}
\usepackage{epsf}
\usepackage{times}
\usepackage{appendix}
\usepackage[normalem]{ulem}
\usepackage{fontawesome}
\usepackage{hyperref}

\usepackage{lineno} 
\newcommand{\pending}[1]{\textcolor{red}{#1}}
\newcommand{\kw}[1]{\textcolor{blue}{#1}}

\DeclareGraphicsExtensions{.jpg,.pdf,.png,.eps,.ps}

\defcitealias{riess18}{R18}

\newcommand{\holicow}{H0LiCOW}
\newcommand{\planck}{{\it Planck}}

\newcommand{\hubbleunit}{km/s/Mpc}
\newcommand{\sqdeg}{deg$^{2}$}
\newcommand{\loglikelihood}{\mathcal{L}}

\newcommand{\be}{\begin{equation}}
\newcommand{\ee}{\end{equation}}
\newcommand{\bea}{\begin{eqnarray}}
\newcommand{\eea}{\end{eqnarray}}

\newcommand{\lcdm}{$\Lambda$CDM}
\newcommand{\LCDM}{$\Lambda$CDM}

\def \five {\textsc{ra5h30dec-55}}

\def \three {\textsc{ra3h30dec-60}}

\def\oz#1{{}}

\def\ell{l}

\newcommand{\comment}[1]{{}}

\def\Davis{1}

\begin{document}

\title{Sounds Discordant:\\ Classical Distance Ladder \& \LCDM -based Determinations of the Cosmological Sound Horizon}

\ifdefined\newapj
\else
\slugcomment{Submitted to \apj}
\fi

\ifdefined\newapj
\author{Kevin Aylor}
\affiliation{Department of Physics, University of California, One Shields Avenue, Davis, CA, USA 95616}
\author{Mackenzie Joy}
\affiliation{Department of Physics and Astronomy, University of Georgia, Sanford Drive, Athens, GA 30602}
\author{Lloyd Knox}
\affiliation{Department of Physics, University of California, One Shields Avenue, Davis, CA, USA 95616}
\author{Marius Millea}
\affiliation{Berkeley Center for Cosmological Physics, LBNL and University of California at Berkeley, Berkeley, California 94720, USA}
\affiliation{Institut dÃ¢ÂÂAstrophysique de Paris (IAP), UMR 7095, CNRS Ã¢ÂÂ UPMC UniversitÃÂ© Paris 6, Sorbonne Universit\'es,
98bis boulevard Arago, F-75014 Paris, France}
\affiliation{Institut Lagrange de Paris (ILP), Sorbonne UniversitÃÂ©s,
98bis boulevard Arago, F-75014 Paris, France}
\author{Srinivasan Raghunathan}
\affiliation{School of Physics, University of Melbourne, Parkville, VIC 3010, Australia}
\affiliation{Department of Physics and Astronomy, University of California, Los Angeles, CA, USA 90095}
\author{W.~L.~Kimmy Wu}
\affiliation{Kavli Institute for Cosmological Physics, University of Chicago, 5640 South Ellis Avenue, Chicago, IL, USA 60637}
\else
\author{Kevin Aylor$^1$, Mackenzie Joy$^2$, Lloyd Knox$^1$, Marius Millea, Srinivasan Raghunathan and W.~L.~Kimmy Wu$^3$}
\altaffiltext{\Davis}{Department of Physics, University of California, One Shields Avenue, Davis, CA, USA 95616}
\altaffiltext{2}{Department of Physics and Astronomy, University of Georgia, Sanford Drive, Athens, GA 30602 }
\altaffiltext{3}{Kavli Institute for Cosmological Physics, University of Chicago, 5640 South Ellis Avenue, Chicago, IL, USA 60637}
\fi

\begin{abstract}
Type Ia Supernovae, calibrated by classical distance ladder  methods, can be used, in conjunction with galaxy survey two-point correlation functions, to empirically determine the size of the sound horizon $r_{\rm s}$. Assumption of the \LCDM\ model, together with data to constrain its parameters, can also be used to determine the size of the sound horizon. Using a variety of cosmic microwave background (CMB) datasets to constrain \LCDM\ parameters, we find the model-based sound horizon to be larger than the empirically-determined one with a statistical significance of between 2 and 3$\sigma$, depending on the dataset. If reconciliation requires a change to the cosmological model, we argue that change is likely to be important in the two decades of scale factor evolution prior to recombination. Future CMB observations will therefore likely be able to test any such adjustments; e.g., 
a third generation CMB survey like SPT-3G can achieve a three-fold improvement in the constraints on $r_{\rm s}$ in the \LCDM\ model extended to allow additional light degrees of freedom. 
\end{abstract}


\ifdefined\newapj
\else
\maketitle
\fi
\section{Introduction}

Classical distance ladder (CDL) approaches using Cepheids and supernovae \citep[hereafter \citetalias{riess18}]{riess18} find higher Hubble constant estimates than those derived from cosmic microwave background (CMB) data, that assume the  standard cosmological model, \LCDM\ \citep{planck18-6}. The statistical significance of this discrepancy has grown over time with fairly steady progress on the distance ladder \citep{riess09,riess11,riess16} and with a sudden jump in precision of the \lcdm\ prediction with the first release of the Planck cosmology data in 2013 \citep{planck13-16}. Comparing the R18 value of \mbox{$H_0 = 73.52 \pm 1.62$ \hubbleunit} with the value inferred from Planck CMB temperature and polarization power spectra plus CMB lensing, assuming \LCDM, \mbox{$H_0 = 67.27 \pm 0.60$ \hubbleunit}, there is a 3.6$\sigma$ discrepancy \citep{planck18-6}.

Along with the reduction of statistical errors in the Cepheids plus supernovae determination of $H_0$, other supporting evidence in favor of a high value of $H_0$ has been growing as well. 
A number of independent analyses of the data have served to largely confirm the conclusions of \citetalias{riess18} \citep{efstathiou14,cardona17,zhang17,feeney18,follin18}. 
\comment{Evidences against the discrepancies in the $H_{0}$ values inferred from \planck{} and CDL measurements also exists.
For example, \citet{wu17} mention that the CDL $H_0$ can depart significantly from the cosmic mean $H_0$ due to the presence of a local void.
However, the authors have also shown that probability of such an occurrence is highly unlikely in a \LCDM{} universe by reporting a sample variance in the \citetalias{riess18} measurement to be 0.3 km/sec/Mpc or 0.2$\sigma$.
\citet{shanks18a} use the ``void''-claim of \citet{wu17} and report a lower value of $H_{0} = 67.4 \pm0.5$ \hubbleunit{} that is highly consistent with \planck{} after accounting for systematic errors in the parallax distances to Cepheids made by the {\it Gaia} satellite \citep{lindegren18}. 
}
Despite claims that the CDL $H_0$ departs significantly from the cosmic mean $H_0$ due to a local void, \citet{wu17} have shown that within \LCDM\ the sample variance in the R18 measurement is only 0.3 km/sec/Mpc, or 0.2$\sigma$.
In addition,
\citet{birrer18} have provided the latest inference of $H_0$ from the \holicow{} \citep{suyu17} collaboration's use of strong-lensing time delays: $H_0 = 72.5^{+2.1}_{-2.3}$ \hubbleunit{}. Combining the result of this completely independent probe of the distance-redshift relation at low redshift, with that from R18, results in a 4.1$\sigma$ discrepancy with the {\it Planck} result quoted above. Finally, the tip of the red giant branch distance method shows consistency with Cepheid distances to a handful of supernova host galaxies \citep{jang17,hatt18a,hatt18b}.

Recently, \citet{shanks18a} claimed that corrections applied to the {\it Gaia} data \citep{lindegren18} in the R18 analysis have introduced significant systematic error, a claim that has sparked a debate \citep{riess18b, shanks18b}. We simply point out here that the \citet{riess16} result of $H_0 = (73.24\pm 1.74)$ \hubbleunit\ makes no use of {\it Gaia} data and is thus immune to this controversy, while nearly as discrepant from the \lcdm\ plus CMB-inferred values.

The case against systematic errors in CMB data as the source of this discrepancy is very strong. First of all, the result from {\it Planck} is robust to choice of frequency channels \citep{planck15-11}, arguing against foreground modeling, or any channel-specific systematic errors as a source of bias in the $H_0$ inference. 
Further, the consistency of \planck{} measurements with WMAP on large angular scales \citep{planck13-1} and the 2500 \sqdeg{} SPT-SZ measurements on small angular scales \citep{hou18,aylor17} also argues against any significant systematic errors on all angular scales relevant for determination of cosmological parameters from \planck\ data.    

Additionally, the conclusion of a low $H_0$ from CMB data and the assumption of \LCDM\ can be reached without any use of {\it Planck} data. The inverse distance ladder results \citep{percival10,heavens14,aubourg14,cuesta15,bernal16,des17,verde17,lemos18,joudaki18,feeney18b} also show that the combination of measurement of the baryon acoustic oscillation (BAO) feature in galaxy surveys, Type Ia supernova observations, and CMB data with or without {\it Planck} (e.g., WMAP9, \citealt{bennett13}) lead to low ({\it Planck}-like) values of $H_0$. 
Finally, \citet{addison18} point out that assuming the \LCDM\ model, using BAO data, and light element abundance measurements as constraints on the baryon-to-photo ratio, one infers a {\it Planck}-like value of $H_0$; i.e., without use of any CMB anisotropy data. 
The above results indicate that systematic errors in CMB data, and \planck\ CMB data in particular, are not the major driver of the discrepancies in inferences of $H_0$.

Recently, some works in the literature have proposed solutions both pre- and post-recombination to address the $H_{0}$ discrepancy. For example, \citet{karwal16, evslin18, poulin18} propose the existence of an early dark energy which reduces the size of the sound horizon subsequently increasing the CMB inferred value of $H_{0}$; \citet{lin18} propose a modified gravity solution at the time of recombination and \citet{chiang18} alter the duration of the recombination to reduce the tension between CMB and CDL results. On the other hand, \citet{divalentino16, divalentino17b, joudaki17} use an extended parameter space and point out that an interacting phantom-like dark energy  with equation of state $w_{\rm DE}<-1$ can also reduce the tension in $H_{0}$ measurements.

In this paper, to further explore what can, and what cannot, explain this discrepancy, we follow 
\citet{bernal16a} and use the sound horizon $r_{\rm s}$, rather than $H_0$, as the point of comparison between CDL and \lcdm-based estimates. With this approach, the BAO data are on the CDL side: Cepheids calibrate Type Ia supernovae, which are then used to determine distances to redshifts for which BAO measurements of the angular size of the sound horizon exist, thereby revealing the size of the sound horizon. We also note that strong-lensing time delays can be used to calibrate the BAO measurements, so we convert the \holicow\ result into a sound horizon constraint. For a more comprehensive use of data sensitive to distances and the expansion rate at low redshifs ($z \lesssim 1$), see \citet[e.g.][]{bernal18}, which includes use of ``cosmic clocks."

\comment{
We find the use of the sound horizon as a point of comparison to be a particularly useful way of examining the data for several reasons. We emphasize one in particular as it bears on another claim of \citet{shanks18a}. The advantage is added insensitivity to extreme changes in the cosmology at $z < 0.1$,  since one does not need to extrapolate to $z=0$. This is helpful because the confusing effect of peculiar velocities complicates inference of the distance-redshift relation at very low redshift. \citet{shanks18a} claim that CDL inferences of $H_0$ are thrown off from the cosmic mean by a local void around us with a radius of about 150 Mpc/$h$, and the associated 500km/sec outflow at $z < 0.1$.  But our CDL inference of $r_{\rm s}$ does not make use of redshift measurements until we get out to the redshifts of the BAO measurements we use, the lowest of which is $z=0.38$. Even granting the claim of a void and its associated outflow (which would be additional strong evidence against \lcdm\ according to \citealt{wu17}), its presence would do nothing to reconcile the $r_{\rm s}$ discrepancies we find.

The remaining advantages are as follows:
first, the \lcdm\ predictions for the sound horizon are more robust than those for $H_0$; second, as with the inverse distance ladder, this approach clarifies that reconciliation can not be delivered by altering cosmology at $z < 1$; third, it serves to clarify that the reconciliation of distance ladder, BAO, and CMB observations via a cosmological solution is likely to include a change to the cosmological model in the two decades of scale factor evolution prior to recombination; and fourth, $\sigma(r_{\rm s})/r_{\rm s}$ from CMB data, assuming \LCDM\ is four times smaller than the $\sigma(H_0)/H_0$ from the same data and assumed model. 
}

We find use of the sound horizon as a point of comparison, to be a particularly useful way of examining the data for several reasons: first, there is added insensitivity with this method to extreme changes in the $z < 0.1$ cosmology since one does not need to extrapolate to $z=0$, circumventing issues with peculiar velocities brought up by ~\citet{shanks18a} ; second,
the \lcdm\ predictions for the sound horizon are more robust than those for $H_0$; third, as with the inverse distance ladder, this approach clarifies that reconciliation can not be delivered by altering cosmology at $z < 1$.
Fourth, it serves to clarify that the reconciliation of distance ladder, BAO, and CMB observations via a cosmological solution is likely to include a change to the cosmological model in the two decades of scale factor evolution prior to recombination; 
and finally, $\sigma(r_{\rm s})/r_{\rm s}$ from CMB data, assuming \LCDM\ is four times smaller than the $\sigma(H_0)/H_0$ from the same data and assumed model.

Since \citet{bernal16a}, we have new supernova data available \citep{scolnic18}, an updated supernova absolute luminosity calibration \citepalias{riess18}, and results from the final data release from {\it Planck} \citep{planck18-6}. We examine the tension in $r_{\rm s}$ given these most recent data. 
As mentioned above, strong-lensing time delays can also be used to calibrate the distance to the BAO redshifts. So we also use the latest \holicow\ results \citep{birrer18} to produce a constraint on $r_{\rm s}$.

We find that the CDL-inferred sound horizons are in 2-3$\sigma$ tension with \lcdm -determined ones\footnote{The inverse distance ladder results in a more significant tension because the BAO error, in this case, gets added to the $r_{\rm s}$ error which is fractionally smaller than the CDL error on $H_0$.  }. While a statistical fluke could explain this discrepancy, we believe there is sufficient evidence to motivate the exploration of cosmological solutions. In Section~\ref{sec:solutions}, we argue that if there is to be a cosmological solution to the discrepancies in $r_{\rm s}$ values, it is likely to be significantly different from \lcdm\ in the two decades of scale factor growth prior to recombination. Such model changes are likely to lead to predictions testable with future CMB data. We examine the $r_{\rm s}$ predictions in the case of a two-parameter extension of the standard cosmological model, and forecasts for $r_{\rm s}$ errors in these extended model spaces given the survey to come from SPT-3G, a third-generation camera outfitted on the South Pole Telescope \citep{benson14,bender18}.

\comment{
We find use of the sound horizon as a point of comparison, to be a particularly useful way of examining the data for several reasons: first, there is added insensitivity with this method to extreme changes in the $z < 0.1$ cosmology since one does not need to extrapolate to $z=0$; second,
the \lcdm\ predictions for the sound horizon are more robust than those for $H_0$; third, as with the inverse distance ladder, this approach clarifies that reconciliation can not be delivered by altering cosmology at $z < 1$.

Since \citet{bernal16a}, we have new supernova data available \citep{scolnic18}, an updated supernova absolute luminosity calibration \citetalias{riess18}, and results from the final data release from {\it Planck} \citep{planck18-6}. We examine the tension in $r_{\rm s}$ given these most recent data. Strong-lensing time delays can also be used to calibrate the distance to the BAO redshifts, so we also use the latest \holicow{} results \citep{birrer18} to produce a constraint on $r_{\rm s}$.

We found that the CDL-inferred sound horizons are in 2-3$\sigma$ tension with \lcdm -determined ones. While a statistical fluke could explain this discrepancy, we believe it is worth exploring potential cosmological solutions. We therefore present a fairly comprehensive discussion of the challenges to finding \lcdm\ model changes that can accommodate the (low) CDL values of the sound horizon. The challenges are considerable due to the agreement of the standard cosmological model with highly precise measurements of CMB power spectra. We discuss extensions that reduce $r_s$ by reducing the sounds speed, and those that do it by reducing the time to recombination -- by either increasing the expansion rate, or by photon cooling. 

Future CMB data may shed light on the matter.We examine the $r_s$ predictions in the case of a two-parameter extension of the standard cosmological model, and forecasts for $r_s$ errors in these extended model spaces given the survey to come from SPT-3G, a third-generation camera outfitted on the South Pole Telescope \citep{benson14}.
}

\section{Models and Data}
\label{sec_model_data}

We present here the empirical, CDL, approach to determining the sound horizon scale, $r_{\rm s}$, and then the more cosmological-model dependent approach. Although the former also requires modeling, we demonstrate with a parametric Spline model for the history of the expansion rate that the results are at most only mildly dependent on cosmological model assumptions. We also describe how we use the recent \holicow\ results \citep{birrer18} to determine $r_{\rm s}$.

\subsection{$r_{\rm s}$ using CDL}
\label{sec_CDL_model}

We describe the formalisms to determine $r_{\rm s}$ empirically in this section. For this purpose, we use the BAO data from the BOSS survey \citep{alam17}, supernova data from the SNeIa Pantheon sample \citep{scolnic18}, and Cepheid data from \citetalias{riess18}.
 
The sound horizon leaves its imprint on the galaxy distribution as a peak in the galaxy two-point correlation function at $r_{\rm s}$, the comoving size of the sound horizon\footnote{More specifically, at the end of the baryon drag epoch; i.e. at $z=z_{\rm drag}$ as defined in \citet{hu96b}. This is often denoted $r_d$, but we use $r_{\rm s}$ to avoid confusion with the diffusion scale.}. 
In redshift space, with galaxy positions recorded in $z$ and angular position on the sky, the sound horizon scale maps into $(\Delta z)_s = H(z)r_{\rm s}$ (the difference in redshift between two galaxies with line-of-sight separation $r_{\rm s}$) and $\theta_s(z) = r_{\rm s}/D_A(z)$ (the angular separation of two galaxies separated by $r_{\rm s}$ perpendicular to the line of sight), where $D_A(z)$ is the comoving angular diameter distance. Thus BAO surveys fundamentally constrain these two quantities and analyses often summarize the constraints as constraints on $H(z)r_{\rm s}$ and $D_A(z)/r_{\rm s}$.  

Type Ia Supernovae are used as ``standardizable" candles, that, after suitable data-dependent corrections, can be reduced to a corrected apparent magnitude with signal modeled by
\be
m^i = M + 5\log_{10}(D^i_L/{\rm Mpc}) + 25
\ee
where $i$ runs over supernovae, the first term on the right is a global, supernova-independent, corrected absolute magnitude, and the 2nd and 3rd terms just follow from the inverse square law for fluxes and the definitions of apparent and absolute magnitudes, with $D_{\rm L}$ the comoving luminosity distance. 

Neglecting, for the moment, the BAO constraints on $H(z)r_{\rm s}$, the BAO and SNe constraints are very similar. Both the BAO and SNe data determine a distance-redshift relationship up to some global scaling factor. In the BAO case the scaling factor is $r_{\rm s}$, and in the supernova case we could take it to be $\ell_{\rm SN} \equiv 10^{-(M+19)/5}$ Mpc\footnote{The choice of ``+19" here is arbitrary; it makes $\ell_{\rm SN} = 1$ Mpc for $M=-19$ which is close to the corrected supernova absolute magnitude.}.  
The two different distances are also very simply related, assuming conservation of photon number, 
via $D_A(z)=D_L(z)/(1+z)$. 

We can relate these observable distance ratios to $H(z)$, assuming negligible mean spatial curvature, via
\bea
\label{eqn:DvZ}
D_A(z)/r_{\rm s} = \frac{c}{r_{\rm s}}\int_0^z\frac{dz}{H(z)} &=& \beta_{\rm BAO}\int_0^z \frac{dz}{\left[H(z)/H_0\right]} \nonumber \\
H(z)r_{\rm s}/c & =&  \frac{1}{\beta_{\rm BAO}} \left[H(z)/H_0\right] \nonumber \\
D_A(z)/\ell_{\rm SN} = \frac{c}{\ell_{\rm SN}} \int_0^z\frac{dz}{H(z)} &=& \beta_{\rm SN}\int_0^z \frac{dz}{\left[H(z)/H_0\right]}
\eea
with $\beta_{\rm BAO} \equiv c/(r_{\rm s} H_0)$ and $\beta_{\rm SN} \equiv c/(\ell_{\rm SN} H_0)$.

Finally, there are the Cepheids. The Supernovae, H0, for the Equation of State of Dark energy (SH0eS) \citep{riess18} program has used Cepheids in 19 different host galaxies with observed SNe Ia to calibrate the mean supernova absolute magnitude. Rather than directly using that calibration, we use the value of $H_0$ that results from the calibration. As one can see in the equations above, the supernova data themselves are sensitive to the combination $\beta_{\rm SN} \equiv c/(\ell_{\rm SN} H_0)$, so specifying $H_0$ allows one to determine $\ell_{\rm SN}$; i.e., it allows for a calibration of the supernova distance measurements.

We use BAO, SNe, and Cepheid data as just (generically) described to infer $r_{\rm s}$, performing our analyses with two different model spaces. One is \lcdm\ with 
\be
H(z)/H_0 = \sqrt{\Omega_{\rm m} (1+z)^3 + \Omega_\Lambda + \rho_\nu(z)/\rho_c + \Omega_\gamma(1+z)^4}
\ee
where
\be
\Omega_\Lambda = 1 - \Omega_{\rm m} - \rho_\nu(z=0)/\rho_c - \Omega_\gamma
\ee
with $\rho_\nu(z)$ calculated for a neutrino background with a temperature today of $T_{\nu,0} = 2.725^\circ$K $(4/11)^{1/3}$, two $m=0$ mass eigenstates and one with mass $m_\nu$ with default value of 0.06 eV and $\Omega_\gamma$ the energy density, in units of the critical density, in a black body of photons with temperature $T_{\gamma,0} = 2.725^\circ$K. The complete set of parameters of this model can be taken to be \{ $\beta_{\rm BAO}, \beta_{\rm SN}, H_0, \Omega_{\rm m}$\}. Note that the sound horizon scale is a derived parameter given by $r_{\rm s} = c/(\beta_{\rm BAO}H_0)$. 

The other model we call the Spline model. For this model, following \citet{bernal16a}, $H(z)/H_0$ is determined by $H(z)$ at 5 locations in $z$ and cubic spline interpolation. The complete set of parameters for the Spline model is \{$\beta_{\rm BAO}, \beta_{\rm SN}, H_0, H_1, H_2, H_3, H_4$\} where $H_i \equiv H(z_i)$ with $z_0 = 0, z_1=0.2, z_2 = 0.57, z_3=0.8$ and $z_3=1.3$, for which we assume a uniform prior over the region with $H(z_i)> 0$. These were the redshift points used by \citet{bernal16a}. We also consider a slightly different choice to check robustness in \S\ref{sec_CDL_spline_based_rs}. 

We note that the Spline model results are not completely free of cosmological assumptions, as the relationship between $H(z)$ and $D_A(z)$ depends on curvature. If it were not for the BAO constraints on $H(z)$ then our Spline model-based inferences of $r_{\rm s}$ would not have any dependence on curvature, as our $H(z)$ parameters can just be thought of, in that case, as a means of parameterizing $D_A(z)$. The reconstructed $H(z)$ would have curvature dependence, but the recovered $r_{\rm s}$ would not. The inclusion of the BAO constraints on $H(z)$ breaks that degeneracy in curvature and brings some dependence of the inferred $r_{\rm s}$ on assumptions about curvature. We will discuss this dependence in the Results \S\ref{sec:results}.

To perform joint analyses of the three datasets, we form a log likelihood $\loglikelihood$ (natural log of the likelihood), given by
\be
\loglikelihood = \loglikelihood_{\rm BAO} + \loglikelihood_{\rm SN} + \loglikelihood_{\rm Cepheids}.
\ee
We now briefly describe each of these likelihoods in turn. 

The log likelihood $\loglikelihood_{\rm BAO}$ has the BAO means and error covariance matrix as described in the BOSS collaboration paper \citet{alam17}, for $D_A(z)/r_{\rm s}$ and $H(z)r_{\rm s}$ at the effective redshifts $z=0.38, 0.51$ and 0.61. These data points are plotted as red squares in Figures \ref{fig:dofz} and \ref{fig:hofz} as constraints on $D_A(z)$ and $H(z)$ given a fiducial value of $r_{\rm s}$. 

We do not include any other BAO data, such as from the 6df galaxy survey \citep{beutler11} and from a BOSS DR12 Ly$\alpha$ absorption cross-correlation analysis \citep{dumasdesbourboux17}. While they provide useful consistency tests of the standard cosmological model they are not as precise as the BOSS galaxy constraints \citep{alam17} and some are also at redshifts greater than the highest redshifts for which we have supernova distance estimates, rendering them uninformative for our main purpose. 

To construct the likelihood for SNe we use the \citet{scolnic18} dataset. They report the redshift-binned estimates of corrected $B$ band SNe apparent magnitudes, corrected to improve the approximation $m = \mu(z_\beta) {\mathbf +} M$ for some global $M$, where $\mu(z_\beta)$ is the distance modulus for redshift $z_\beta$. We thus model the data as
\be
\label{eqn:apparentmagnitudemodel}
m(z_\beta) = M + 5 \log_{10}(D_{\rm L}(z_\beta)/{\rm Mpc}) + 25.
\ee
The absolute magnitude $M$ is the more usual way of specifying the calibration of the supernovae. Taking $\ell_{\rm SN}$ introduced above to have the fiducial value of 1 Mpc for a fiducial value of $M=-19.3$ (for specificity) then Eq.~\ref{eqn:apparentmagnitudemodel} can be rewritten to swap in $\ell_{\rm SN}$ for $M$:
\be
m(z_\beta) = 5 \log_{10}\left(\frac{D_{\rm L}(z_\beta)}{\ell_{\rm SN}} \right) + 5.7.
\ee
We form a likelihood that is Gaussian in the apparent magnitudes, with covariance matrices that include the statistical and systematic errors as reported in \citet{scolnic18}. The data points are plotted as green circles in Fig. \ref{fig:dofz} as constraints on $D_A(z) = D_L(z)/(1+z)$ for a fiducial value of $M$.

For the ``Cepheids'' log likelihood, we take 
\be
\loglikelihood_{\rm Cepheids} = -\frac{\left(73.52 - H_0\right)^2}{2\times 1.62^2}
\ee
where $H_0$ is our model Hubble constant in \hubbleunit{} and the numbers in the likelihood are from 
\citetalias{riess18} measurement $H_0 = 73.52 \pm 1.62$ \hubbleunit. Note that, just like for $r_{\rm s}$ with the BAO data, $\ell_{\rm SN}$ is a derived parameter given by $c/(\beta_{\rm SN}H_0)$. The supernova absolute magnitude parameter $M$ can likewise be derived from $\ell_{\rm SN}$.

\subsection{$r_{\rm s}$ and strong-lensing time delays}
\label{sec_sltd}
We also consider strong-lensing time delay (SLTD) data \citep{birrer18} as a means of calibrating the BAO. A given SLTD system is sensitive to the ratio $D_{\rm d}D_{\rm s}/D_{\rm ds}$ where $D_{\rm d}$ is the distance to the lens (typically near $z\,{\sim}\,0.5$), $D_{\rm s}$ the distance to a lensed quasar (typically near $z\,{\sim}\,1.5$), and $D_{\rm ds}$ the distance between the two. This quantity is inversely proportional to $H_0$, and, in \LCDM, its dependence on the exact shape of $H(z)$ (given largely by $\Omega_m$) is weak enough to, even with very weak priors on the matter density, produce a strong constraint on $H_0$. We can thus use this constraint to anchor the BAO point instead of the Cepheids without any other additional external data. In practice, we simply combine the constraint on $\beta_{\rm BAO}$ which we get from SNe and BAO with the $H_0$ reported by \cite{birrer18}, propagating Gaussian error bars in quadrature.

We note that this analysis is approximate because we have not jointly analyzed the datasets; improved constraints on the matter density from the SNe+BAO data could further tighten the \holicow\ result. However, this effect is likely to be small given the weak dependence on the $\Omega_{\rm m}$ prior reported by \cite{birrer18}. Note that this analysis does assume \LCDM, in particular that the shape of $H(z)$ follows the expectation from \LCDM\ between today and the quasar redshifts of $z\,{\sim}\,1.5$. While the SNe strongly constrain the shape at somewhat lower redshifts, there is, at least in theory, the possibility that the \holicow\ inference of $H_0$ and thus our corresponding $r_{\rm s}$ inference could be somewhat thrown off by a change to $H(z)$ right around $z\,{\sim}\,1.5$. We have not attempted a joint spline fit of SNe+BAO+\holicow, but such a test could reveal to what extent this is a possibility (although, of course, the \holicow\ and Cepheid determinations of $H_0$ are already in good agreement, arguing against this possibility).

\subsection{$r_{\rm s}$ from \LCDM\ plus CMB data}
\label{sec_CMB_model}
We have just reviewed how one can infer $r_{\rm s}$ in an empirical manner using the CDL. Here we describe how one can adopt a model and directly calculate $r_{\rm s}$. The comoving size of the sound horizon is given by 
\be
r_{\rm s} = \int_0^{t_{\rm d}} c_s(a) dt/a(t) = \int_0^{a_d} da \frac{c_s(a)}{a^2 H(a)}
\label{eqn:soundhorizon}
\ee
where $c_s(a)$ is the sound speed as a function of the scale factor and $a_{\rm d}$ is the scale factor at the end of the baryon drag epoch. In the \LCDM\ model $r_{\rm s}$ is completely determined by the baryon-to-photon ratio, for its influence on $a_{\rm d}$ and $c_s(a)$, and the matter density $\omega_{\rm m} \equiv \Omega_{\rm m}h^2$ for its influence on $a_{\rm d}$ and $H(a)$. With these parameters constrained, or any other relevant parameters there might be in extended model spaces, one can then calculate a constraint on $r_{\rm s}$.

We use the CMB datasets from the Atacama Cosmology Telescope \citep[ACTPol,][]{louis16}, \planck\ \citep{planck18-6}, the South Pole Telescope: SPT-SZ \citep{aylor17} and SPTpol \citep{henning18}, and the Wilkinson Microwave Anisotropy Probe (WMAP, \citealt{bennett13}). We look at subsets of the Planck data as well. Significant constraints on $r_{\rm s}$ come from each of the three dominant power spectra: $C_l^{TT}, C_l^{TE}$, and $C_l^{EE}$, as well as from $C_l^{TT}$ at $\ell < 800$ and $C_l^{TT}$ at $\ell > 800$.

\section{Results and Discussion}
\label{sec:results}

\begin{figure}[!tbh]
\includegraphics[width = .48\textwidth]{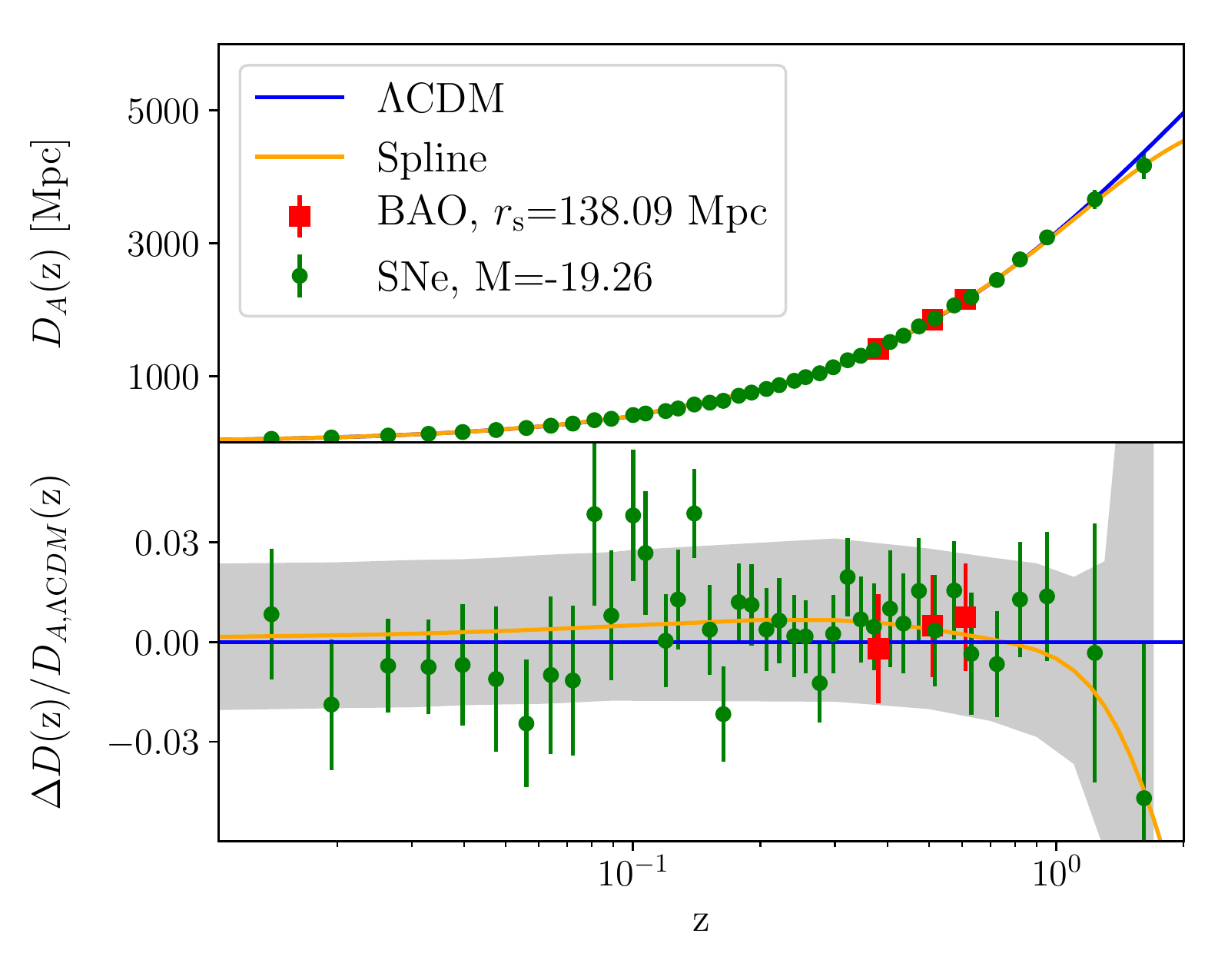}
\caption{Comoving angular-diameter distance measurements, $D_A(z)$, together with best-fit models. BAO results have been converted from $D_A(z)/r_{\rm s}$ to $D_A(z)$ by assumption of $r_{\rm s} = 138.09$ Mpc. Supernovae distance moduli have been converted to $D_A(z)$ assuming $M=-19.26$. In the residuals panel, $\Delta D_A(z) = D_A(z)-D_{A,\Lambda{\rm CDM}}(z)$ where $D_{A,\Lambda{\rm CDM}}(z)$ is the comoving angular-diameter distance for the best-fit $\Lambda$CDM cosmology. The gray band shows the 68\% confidence interval 
for the spline model.}
\label{fig:dofz}
\end{figure}

\begin{figure}[!tbh]
\includegraphics[width = .48\textwidth]{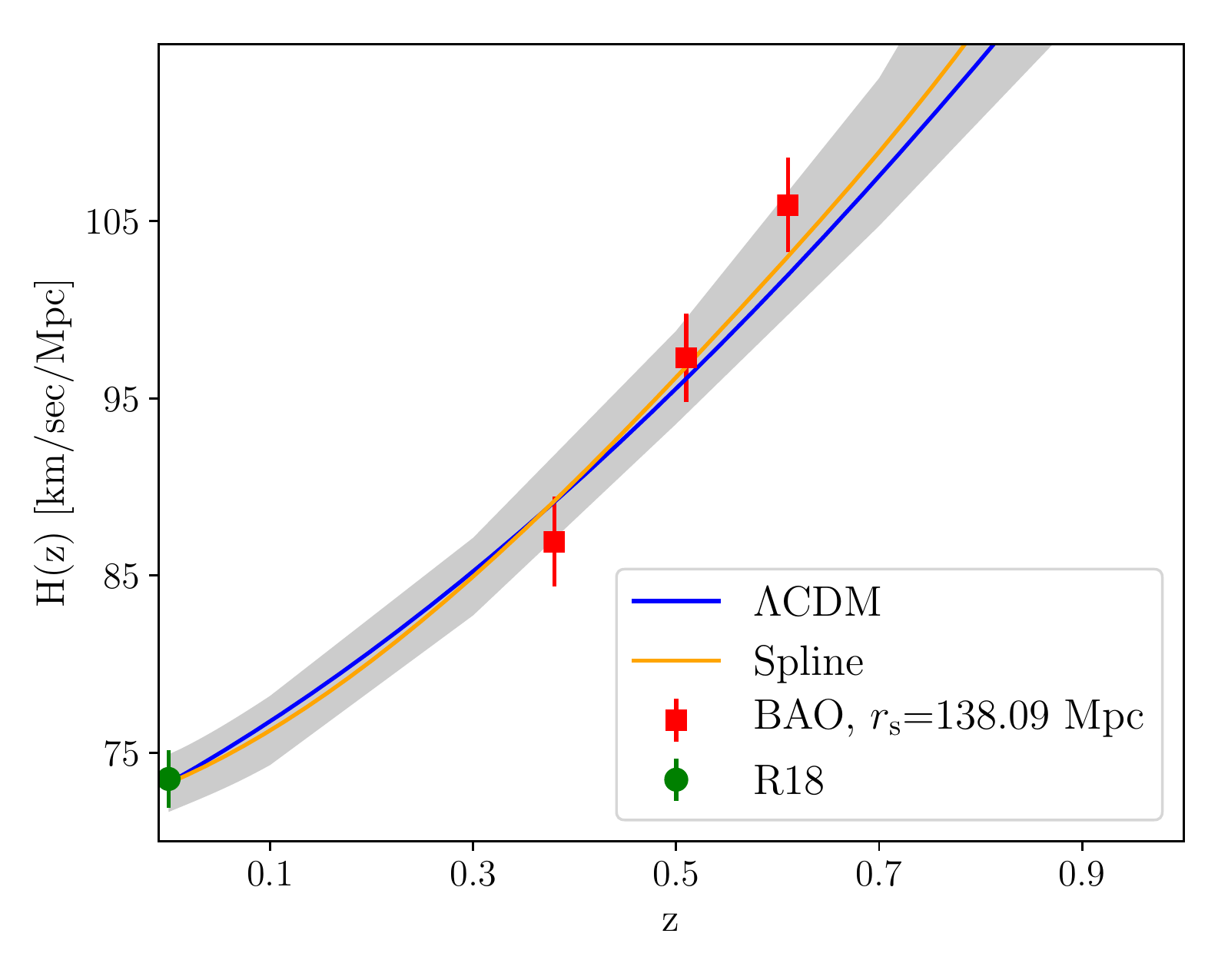}
\caption{Expansion rate measurements together with best-fit models. BAO data have been converted to $H(z)$  by assumption of $r_{\rm s} = 138.09$ Mpc. The gray band shows the 68\% confidence interval 
for the spline model.}
\label{fig:hofz}
\end{figure}

Before presenting the constraints on $r_{\rm s}$ from different methods described in \S\ref{sec_model_data}, in Figures \ref{fig:dofz} and \ref{fig:hofz} we display the BAO and SNe data as they constrain $D_A(z)$ and $H(z)$.
For these figures, we assume the fiducial values $r_{\rm s}=138.09$ Mpc and $M=-19.26$, which are the best-fit values for the Spline model parameter space described in \S\ref{sec_CDL_spline_based_rs} given the BAO, SNe, and Cepheid data. We choose the best-fit values from the spline model, as opposed to the \lcdm\ model, primarily for specificity, and secondarily in order to have less model dependence in the resulting distance estimates.

Examining the residuals from these fits in Fig.~\ref{fig:dofz} and Fig.~\ref{fig:hofz} we see no obvious problems for either the \LCDM\ model or the Spline model. 
For the \LCDM\ model, we find for the best fit $\chi^2_{\rm SNe} = 39.3$ and $\chi^2_{\rm BAO}$ = 3.5, summing to $\chi^{2}_{\rm tot}$ = 42.8 for 43 degrees of freedom (40 SNe data points, 6 BAO data points, and 3 parameters, not counting $H_0$). 
For the Spline model, we find the best fit $\chi^2_{\rm SNe} = 38.0$ and $\chi^2_{\rm BAO}$ = 3.4, with $\chi^{2}_{\rm tot}$ = 41.4 for 40 degrees of freedom (6 parameters, once again not counting $H_0$). 

\begin{figure*}[!tbh]
\begin{center}
\includegraphics[width=.9\textwidth, keepaspectratio]{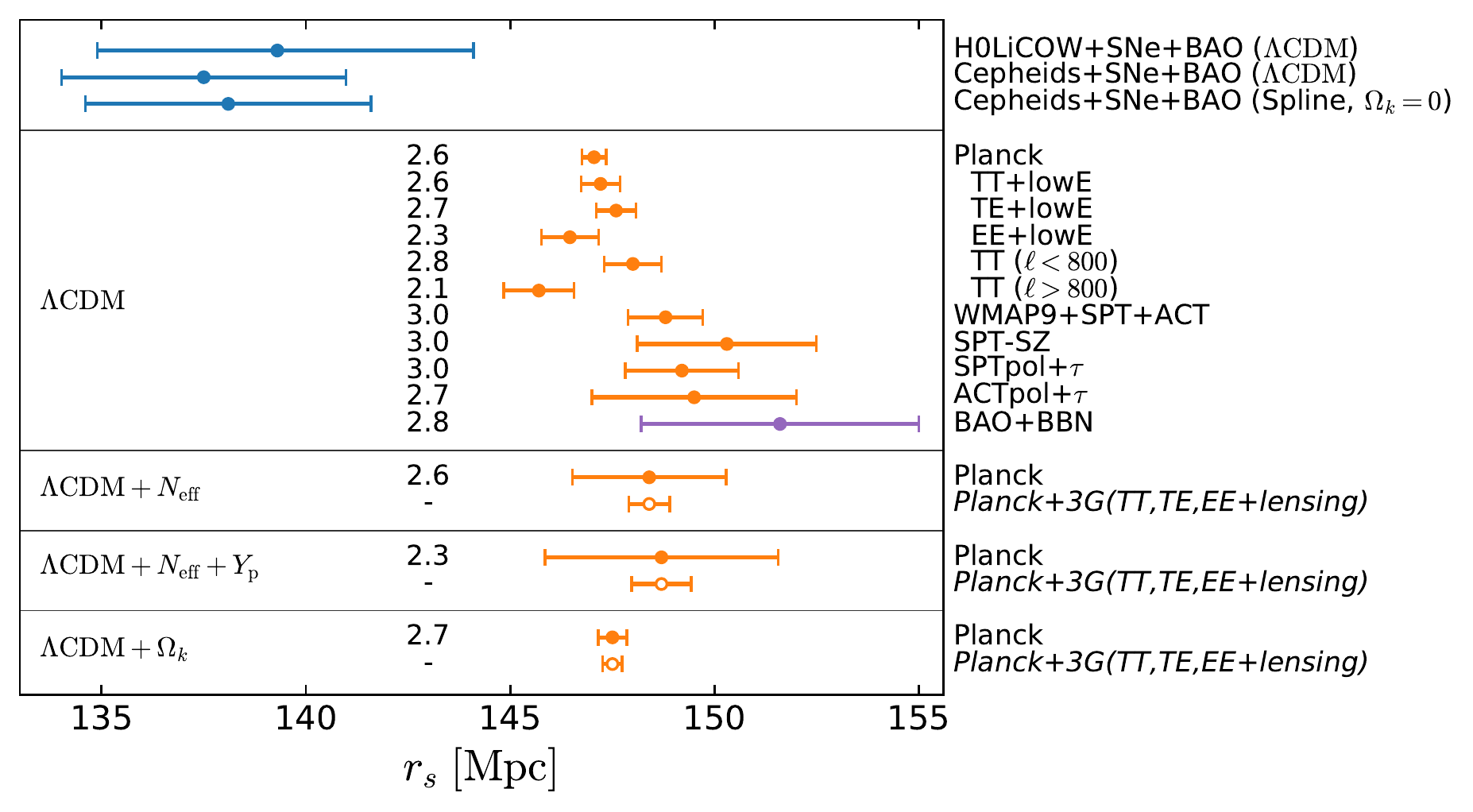}
\end{center}
\caption{Sound horizon determinations from existing data (solid symbols) and forecasts (open symbols). The numbers down the middle give the difference with the Cepheids+SNe+BAO Spline model result for $r_{\rm s}$ in units of the standard deviation, with the standard deviation computed via quadrature sum. We see that the classical distance ladder constraints (top panel) on $r_{\rm s}$ come out systematically lower than the \lcdm-based constraints (biggest panel). The three model extensions considered in the three remaining panels do not significantly weaken the discrepancy. Code and data for this figure is available here:  
\href{https://github.com/marius311/sounds\_discordant\_plot/blob/2c4735a17229ec14f3e54e2a594803d2a1bb34ca/summaryplot.ipynb}{\faGithub}
}
\label{fig:Summary}
\end{figure*}

\comment{
In this section, we first infer $r_{\rm s}$ using the CDL approach from the \LCDM\ and the Spline model, and find negligible dependence on the assumed model. We then compare these CDL results to the model-based $r_{\rm s}$ inference from CMB data, and find them to be discrepant at the $2$ to $3\sigma$ level. Neither of these approaches is sensitive to assumptions about the cosmological model after recombination. Therefore, if the origin of the discrepancies is cosmological, the cosmological solution must make its important changes at times prior to recombination.

In Fig.~\ref{fig:Summary} we compare constraints on $r_{\rm s}$ arising from both the more-empirical CDL method, and the much more model-dependent method. For the model-dependent method, for two extensions of the \LCDM\ model, we also provide forecasts for uncertainties expected from the combination of \planck{} and a stage-3 CMB survey: SPT-3G \citep{benson14}.
}

Now we turn to results reporting the inferred value of $r_{\rm s}$ using the CDL approach from the \LCDM\ and the Spline model in \S\ref{sec_CDL_based_rs}. 
This is followed by the results obtained using CMB data for the \lcdm{} model in \S\ref{sec_LCDM_CMB_based_rs}. 
Next, we discuss the \mbox{$2$ to $3\sigma$} tension in the value of $r_{\rm s}$ obtained from these two methods in \S\ref{sec_tension_CDL_CMB}.
In \S\ref{sec_forecast}, we look at a couple model extensions and forecast the expected constraints on $r_{\rm s}$ that can be obtained by combining \planck{} results with SPT-3G \citep{benson14}, a stage-3 CMB temperature and polarization survey. 
Finally, in \S\ref{sec:solutions},we argue that if the origin of the discrepancies is cosmological, the cosmological solution must make its important changes at times prior to recombination.

\subsection{CDL based constraints}
\label{sec_CDL_based_rs}
We begin our discussion with our first result from a combination of the $H_0$ constraint 
(that we refer to as ``Cepheids", \citetalias{riess18}), used for calibrating the Pantheon binned distance moduli (``SNe", \citealt{scolnic18}), which in turn are used to calibrate the BAO distance and $H(z)$ constraints from BOSS galaxies (``BAO", \citealt{alam17}). 
The CDL based $r_{\rm s}$ results are shown as blue circles in the top panel of Fig.~\ref{fig:Summary}.

\subsubsection{CDL + \lcdm{}
\label{sec_CDL_lcdm_based_rs}}
First, we have assumed the \LCDM\ model -- using it to provide the parameterized shape of $H(z)/H_0$. We find 
\be
r_{\rm s} = (137.6 \pm 3.45)\  {\rm Mpc}.
\label{eqn:CDL_LCDM_result}
\ee

As a point of comparison we mention a result from \citet{addison18}. 
They take a more comprehensive set of BAO data, including constraints at lower redshift from galaxy surveys \citep{beutler11, ross15},  
and higher redshift constraints from BOSS Lyman-$\alpha$ \citep{font-ribera14, delubac15, bautista17} and find, from the BAO data themselves, assuming the \LCDM\ model, that
$ H_0 r_{\rm s} = (10119 \pm 138)$ km/sec.
Combining this with the \citetalias{riess18} 
result for $H_0$ it becomes
\begin{equation}
r_{\rm s} = (137.7 \pm 3.7)\ {\rm Mpc}
\ee
This result is nearly the same, in mean and standard deviation, as our own CDL + \LCDM\ result. The lack of reduction in uncertainty, despite the much greater amount of BAO data, is due in part to the lack of use of the SNeIa data\pending{\sout{,}} which increases uncertainty in $\Omega_{\rm m}$, and therefore the shape of $D_A(z)$. The other important factor in the lack of reduction is that the BOSS galaxy data are unmatched in precision.
 
Our second CDL + \LCDM\ result comes from replacing Cepheids \citepalias{riess18} with the SLTD data from H0LiCOW \citep{birrer18} like explained in \S\ref{sec_sltd}.
From our SNeIa + BAO data we have 
$\beta_{\rm BAO} \equiv c/(r_{\rm s} H_0) = 29.7 \pm 0.37$. Combining this with $H_0 = 72.5^{+2.1}_{-2.3}$ \hubbleunit{} from \citet{birrer18} we find
\be
r_{\rm s} = 139.3^{+4.8}_{-4.4}\ {\rm Mpc}.
\ee
That uncalibrated supernovae, combined with BAO data, put a strong constraint on the product $r_{\rm s}H_0 (=c/\beta_{\rm BAO})$ was previously mentioned in \citet{verde17}.

\subsubsection{CDL + Spline}
\label{sec_CDL_spline_based_rs}

To explore the model-dependence of the CDL method for $r_{\rm s}$ inference, we now drop the assumption of \LCDM\ for parameterization of the shape of $H(z)/H_0$ and replace it with our Spline model. Because our BAO results span such a small range of redshift, we can expect that there is very little sensitivity of the inferred $r_{\rm s}$ to the choice of parameterization, as long as it is not varying rapidly on redshift intervals comparable to the redshift span of the BAO measurements. With the four-parameter model described in the previous section we indeed find a very similar result to the \LCDM\ result:
\be
r_{\rm s} = (138.0 \pm 3.59)\ {\rm Mpc}.
\label{rs_from_cdl_spline}
\ee

That this sound horizon result is a little bit larger is consistent with what we see in the residuals panel of Fig.~\ref{fig:dofz}. Namely, the SNe data largely sit above the \LCDM\ best-fit curve in the redshift interval with the BAO data. The increased freedom of the empirical model reduces the influence of the SNe outside of this redshift range, boosting $D(z)$ in this interval with the result that $r_{\rm s}$ is slightly larger. Note though that statistically, this is a very small shift of less than 0.2$\sigma$. 

More importantly, because the \LCDM\ and Spline results for $r_{\rm s}$ are basically the same, including in the uncertainty, we can conclude that the CDL sound-horizon determination is highly model independent. In particular, it is, at most, very weakly dependent on any assumptions about the shape of the distance-redshift relationship. As a further check, we performed an analysis with Spline points moved to $z = $\{0, 0.2, 0.5, 0.8, 1.1\} away from our baseline $z=\{0, 0.2, 0.57, 0.8, 1.3\}$ (see \S\ref{sec_CDL_model}) and obtain $r_{\rm s} = 137.7 \pm 3.60$ Mpc indicating that our results are not highly sensitive to the choice of pivotal redshift points.

Before closing this subsection we comment on the dependence of the CDL result for $r_{\rm s}$ on curvature. Using R16 for the $H_0$ constraint, \citet{betoule14} for the SNeIa data, and the same BOSS BAO data, \citet{verde17} found, also for a phenomenological parameterization of $H(z)$, that $r_{\rm s} = 138.5\pm 4.3$ Mpc assuming $\Omega_k = 0$. This is consistent with our result to within 0.2$\sigma$. When they marginalize over $\Omega_k$ they find $r_{\rm s} = 140.8\pm 4.9$ Mpc. This is a $\sim 0.5 \sigma$ shift, which indicates that were we to relax our zero curvature assumption, it might have some impact on the significance of our results. We caution against seeing this small shift as possibly leading to a resolution between the CMB and CDL data. To get the full magnitude of this shift requires the curvature to be quite far from zero. The \citet{verde17} constraint on $\Omega_k$ in this analysis is $\Omega_k = 0.49 \pm 0.64$.
Such a large value of $\Omega_k$ is highly disfavored by CMB data; in the \lcdm + $\Omega_k$ model the Planck temperature and polarization power spectra lead to $\Omega_k = -0.044 \pm 0.034$.

\subsection{\LCDM-based constraints with and without CMB data}
\label{sec_LCDM_CMB_based_rs}
We now turn to the model-based determinations of the sound horizon, focusing first on the \LCDM\ model results. 
To examine robustness of sound-horizon determination we show results for many choices of CMB datasets (orange circles in the biggest panel of Fig.~\ref{fig:Summary}).
We see some scatter in these inferences of $r_{\rm s}$, with all of them between 2 and 3$\sigma$ larger than the spline-based CDL result.

A curious feature of the scatter in the \LCDM\ results is that those datasets that lead to lower values of $H_0$, such as using \planck{} temperature power spectrum (TT) data restricted to $\ell > 800$ (+ lowE), which are thus {\em more} discrepant with the CDL value of $H_0$, also lead to values of $r_{\rm s}$ that are {\em less} discrepant with the CDL, and vice versa. This pattern can be understood as follows. First, recall that the comoving size of the sound horizon is given by Eq.~\ref{eqn:soundhorizon}, which, in the \LCDM\ model depends only on the baryon-to-photon ratio and the matter density $\omega_{\rm m}$.
The fluctuations in \LCDM -based $r_{\rm s}$ inferences from CMB data are almost entirely driven by fluctuations in $\omega_{\rm m}$. The short explanation for the positive correlation between $H_0$ and $r_{\rm s}$ fluctuations is that upward fluctuations drive $r_{\rm s}$ downward, and $H_0$ downward as well.

The positive correlation between $r_{\rm s}$ and $H_0$ can be understood as follows. If the radiation density were completely negligible for the calculation of the sound horizon, then, from the Friedmann equation, $\delta H(a)/H(a) \propto 0.5 \delta \omega_{\rm m}/\omega_{\rm m}$ so we have $\delta r_{\rm s}/r_{\rm s} \propto -0.5 \delta \omega_{\rm m}/\omega_{\rm m}$. The radiation softens this response to closer to $\delta r_{\rm s}/r_{\rm s} \propto -0.25 \delta \omega_{\rm m}/\omega_{\rm m}$ \citep{hu01a}. To keep the angular size of the sound horizon fixed (in order to stay at high CMB data likelihood), we have for the distance from here to $z=z_{\rm d}$, $\delta D/D = \delta r_{\rm s}/r_{\rm s} = -0.25 \delta \omega_{\rm m}/\omega_{\rm m}$. For the model to achieve this softened response of the distance to the matter density (softened to a $-0.25$ exponent as opposed to $-0.5$) there has to be a fluctuation in the dark energy density that is anti-correlated with the matter density fluctuation, with the result that $\delta H_0/H_0$ has the same sign from $\delta r_{\rm s}/r_{\rm s}$, as also explained in \citet{hou14}.  
Perhaps of particular note regarding this positive correlation between $r_{\rm s}$ and $H_0$ fluctuations, is that those datasets that are somewhat more consistent with the CDL for $H_0$ than is the case for {\it Planck}, are {\em less} consistent with the CDL for $r_{\rm s}$. This is the case for \planck{} TT ($\ell < 800$), WMAP9+SPT+ACT~\citep{calabrese17}, SPT-SZ~\citep{aylor17}, and SPTpol~\citep{henning18}. These fluctuations toward higher $H_0$, if they go far enough to reconcile with \citetalias{riess18}, 
end up being discrepant with BAO data (given the \LCDM\ model) as noted in \citet{hou14}. 

While the above results indicate robustness to the choice of the CMB data, \citet{addison18} have demonstrated that $r_{\rm s}$ can be estimated, assuming \LCDM, without any CMB anisotropy data at all. They  use a combination of BAO data and constraints on $\omega_b$ from inferences of the primordial abundance of deuterium relative to hydrogen (D/H) \citep{cooke16}. 
Within \LCDM\, $r_{\rm s}$ is entirely determined by $\omega_m$ and $\omega_b$ via Eq.~\ref{eqn:soundhorizon}.
Given the assumption of \LCDM\ the BAO data can be used to constrain $\omega_m$. This constraining power arises from the degeneracy breaking power of separately parallel and perpendicular constraints at several different redshifts. The primordial D/H ratio resulting from big bang nucleosynthesis (BBN) is highly sensitive to the baryon-to-photon ratio and can therefore be used to estimate $\omega_b$.
 \citet{addison18} combine galaxy and Lyman-$\alpha$ forest BAO with a precise estimate of the primordial deuterium abundance \citep{cooke16} to find $r_{\rm s} = 151.6 \pm 3.4 $ Mpc.
The BAO+BBN based $r_{\rm s}$ is shown in Fig.~\ref{fig:Summary} in purple (rather than orange like CMB) as it relies on BAO data that has also been used for the CDL determination, and whose interpretation is dependent in this case on late-time assumptions of the \LCDM\ model. 
This result, like the CMB-based estimates, is also discrepant with the CDL measured values of $r_{\rm s}$.

\comment{
Finally, instead of using CMB data to determine the relevant \LCDM\ parameters for calculating $r_{\rm s}$ via Eq.~\ref{eqn:soundhorizon}, one can use a combination of BAO data and constraints on $\omega_b$ from BBN measurements \citep{addison18}. Within \LCDM\, $r_{\rm s}$ is entirely determined by $\omega_m$ and $\omega_b$. Given the assumption of \LCDM\ the BAO data are able to constrain $\omega_m$ (in particular, due to the degeneracy breaking power of using separately parallel and perpendicular constraints, and at several different redshifts), and the BBN data are able to constrain $\omega_b$. For our $r_{\rm s}$ constraints, we use directly the numbers given by \cite{addison18} for the BAO+D/H case with the theoretically-determined reaction rates \citep{cooke16}.

We have shown robustness to choice of CMB data. 
Now we will explore other analyses in the literature that determine $r_{\rm s}$ without including CMB data.
\citet{addison18} and \citet{des18} both estimate $H_0$ in the \LCDM\ model without use of any CMB data. \citet{addison18} combine galaxy and Lyman-$\alpha$ forest BAO with a precise estimate of the primordial deuterium abundance to find $H_0=66.98\pm 1.18$ \hubbleunit. 
\citet{des18} combine Dark Energy Survey Year-1 clustering and weak lensing data with BAO and BBN constraints on the baryon-to-photon ratio to find, assuming the \LCDM\ model, that $H_0 = 67.2^{+1.2}_{-1.0}$ \hubbleunit.

\citet{addison18} also report, from the same set of data they use for the Hubble constant constraint above, a sound horizon of $r_{\rm s} = 151.6 \pm 3.4 $ Mpc.
We include this result in our summary figure, Fig.~\ref{fig:Summary} but color it differently as it relies on BAO data that has also been used for the CDL determination, and whose interpretation is dependent in this case on late-time assumptions of the \LCDM\ model. 
The result has some dependence on the $d(p,\gamma)^3$He reaction rate. \sout{The result we quote here,} \kw{For the result we quote here} and place in Fig.~\ref{fig:Summary}, \kw{we} assume the theoretically-determined rate. \kw{If we use} \sout{Use of} the empirically calculated rate, \kw{we get} \sout{leads to} $r_{\rm s} = 149.2\pm 3.6$ Mpc.
}

\subsection{Tension in $r_{\rm s}$}
\label{sec_tension_CDL_CMB}
The tension between these two means of inferring $r_{\rm s}$, the CDL measurement vs. the \LCDM\ calculation, is the main result of this paper. Cast in terms of $r_{\rm s}$, rather than in terms of $H_0$, it is clear -- as the inverse distance ladder approach also suggests -- that if the solution to the discrepancies lies in cosmology, we need modifications to cosmology at early times, not late times. 
We need a model that, given the CMB data, produces a smaller sound horizon. 
We discuss this further in \S\ref{sec:solutions}.

\comment{
We also consider here an extension that allows for additional non-interacting, relativistic species; i.e., ones that behave just like massless neutrinos (the \LCDM + $N_{\rm eff}$ model space) and an extension to that in which the primordial fraction of baryonic mass in Helium, $Y_{\rm P}$, is also allowed to vary freely (the \LCDM + $N_{\rm eff} + Y_{\rm P}$ model space). We see that these extensions do very little, if anything, to relieve the tension with the CDL result. We will discuss these results further in the next section.
}

\subsection{Extensions and Forecasts}
\label{sec_forecast}

An extension of \lcdm\ often considered for its possibility of reducing $H_0$ tension is to let the effective number of light and non-interacting degrees of freedom, $N_{\rm eff}$, be a variable, freed from its \lcdm\ value of 3.046. One of the hindrances to adjustment of $N_{\rm eff}$ is that it leads to a change in the ratio of sound horizon to damping scales \citep{hu96a,bashinsky04,hou13}, a change that is not preferred by CMB data. To loosen up these damping-scale constraints, we also consider allowing the primordial fraction of baryonic mass in Helium, $Y_{\rm P}$, to be freed from its BBN-consistent value. We see that these extensions do very little, if anything, to relieve the tension with the CDL result. They do increase the uncertainty significantly in $r_{\rm s}$, but the uncertainty remains sub-dominant to the CDL uncertainty, so there is not much impact on the significance of the difference. 

To give an example of robustness to changes to late-time cosmology, we also show results for the extension to free mean curvature, \lcdm\ + $\Omega_{\rm K}$. As expected, allowing curvature to float has very little impact, if any, on the inference of $r_{\rm s}$.

Next we forecast the expected constraints on $r_{\rm s}$ to come from a combination of \planck{} and the final results of the SPT-3G survey that is currently underway.
The constraints are presented in Fig. \ref{fig:Summary} as open circles.
 
In the \LCDM\ + $N_{\rm eff}$ model space the error in $N_{\rm eff}$ will reduce by a factor of 2 compared to \planck-only results. 
The resulting reduction in $r_{\rm s}$ follows from this $\sigma(N_{\rm eff})$ reduction plus reduction in the matter density uncertainty as well. In the \LCDM\ + $N_{\rm eff} + Y_{\rm P}$ model space the area of the 68\% confidence region is reduced by a factor of 2.8 with the inclusion of SPT-3G compared to \planck{} alone. Because there is no degeneracy between $\Omega_{\rm K}$ and $r_{\rm s}$ the improvement of constraint on $r_{\rm s}$ in the \lcdm\ + $\Omega_{K}$ model space is less dramatic. 

We also see that constraints from current CMB data on $r_{\rm s}$ do not change much with the extension from \lcdm\ to \lcdm +$\Omega_K$. This is expected as the inference is not sensitive to the distance to the last-scattering surface. This insensitivity to late-time physics was previously noted by \citet{verde17b}.

\comment{
These extensions do increase the uncertainty significantly in $r_{\rm s}$, but the uncertainty remains subdominant to the CDL uncertainty, so there is not much impact on significance of difference. 
We have also investigated to what extent the uncertainty in the extended model spaces can be reduced by near-future observations. In particular, we have made forecasts for errors on $r_{\rm s}$ to come from the combination of \planck{} and the SPT-3G survey. The assumptions that go into the forecasts are presented in the Appendix.

One can see in Fig.~\ref{fig:Summary} that the addition of SPT-3G data will greatly shrink the uncertainty in the $r_{\rm s}$ errors in the \LCDM\ + $N_{\rm eff}$ and \LCDM\ + $N_{\rm eff}$ + $Y_{\rm P}$ spaces. Constraints on these parameters will tighten as well. In the \LCDM\ + $N_{\rm eff}$ model space the error in $N_{\rm eff}$ will reduce by a factor of 2 from using \planck{} to using \planck{} + SPT-3G. The resulting reduction in $r_{\rm s}$ follows from this $\sigma(N_{\rm eff})$ reduction plus reduction in the matter density uncertainty as well. In the \LCDM\ + $N_{\rm eff} + Y_{\rm P}$ model space the area of the 68\% confidence region is reduced by a factor of 2.8 with the inclusion of SPT-3G compared to from \planck{} alone. 
}

\section{Cosmological Solutions}
\label{sec:solutions}

The inverse distance ladder papers we cited earlier, and also \citet{poulin18}, indicate that the combination of BAO and supernova data make a cosmological solution unlikely with changes restricted to $z \la 1$. Here we go further and argue that any viable cosmological solution to sound horizon discrepancies is likely to differ significantly from the standard cosmological model in the two decades of scale factor expansion immediately prior to recombination. Changes that are only important earlier cannot reduce the sound horizon significantly.
This is because in the standard cosmological model, near the best-fit location in parameter space given \planck\ data, greater than 95\% of $r_{\rm s}$ is generated in the final two decades of scale factor growth prior to recombination. 

What about changes after recombination? These would have to make a fractional change in $r_{\rm s}$ of $\delta r_{\rm s}/r_{\rm s} = x$ where $x \simeq -0.07$ to bring the model $r_{\rm s}$ values in line with the CDL values. If the changes were only important after recombination, then our $r_{\rm s}$ calculation is unchanged so we still have $\delta r_{\rm s}/r_{\rm s} \simeq -0.25 \delta \omega_{\rm m}/\omega_{\rm m}$ so we need $\delta \omega_{\rm m}/\omega_{\rm m} = -4x$. To preserve $\theta_{\rm s}$ (which we would need to do to stay at high likelihood given the CMB data; e.g., \citealt{pan16}) we would also need to change the angular-diameter distance to last-scattering by $\delta D/D = x$. However, another important length scale for interpretation of CMB data, the comoving
 size of the horizon at matter-radiation equality, $r_{\rm EQ} = c/(a_{\rm EQ}H_{\rm EQ})$, responds much more rapidly to changes in $\omega_{\rm m}$. We find, assuming \lcdm, as is appropriate here, $\delta r_{\rm EQ}/r_{\rm EQ} = -3 \delta \omega_{\rm m}/\omega_{\rm m} = 12x$ and therefore the change in distance required to keep $\theta_{\rm EQ} = r_{\rm EQ}/D$ from changing would be 12 times greater than required to keep $\theta_{\rm s}$ fixed. We can not make changes to the late-time cosmology, and therefore $D$, that keep both of these angular scales fixed. To make this work, the changes in the post-recombination cosmology would have to introduce new anisotropies that would confuse our inference of $\theta_{\rm EQ}$ and/or $\theta_{\rm s}$. The consistency of the \lcdm\ results for $r_{\rm s}$ (which depend primarily on $\omega_{\rm m}$, which is inferred from $\theta_{\rm EQ}$; see, e.g., Section 4 of \citet{planck16-51} ) across angular scale argue against this possibility. We find it to be highly unlikely that whatever confuses our interpretation of the $\ell < 800$ TT data (perhaps ISW effects) would also similarly confuse our interpretation of the $\ell > 800$ TT data, as well as our interpretation of other data selections such as TE+lowE. 
 
Our claim in this section, that any viable cosmological solution is likely to include significant changes from \lcdm\ in the epoch immediately prior to recombination, is an interesting one, as this is an epoch that we will probe better with improved measurements of CMB polarization (and also temperature on small angular scales). It has this exciting implication: viable cosmological solutions are likely to make predictions that are testable by so-called stage 3 CMB experiments, as well as CMB-S4. 

Soon after we posted this paper on the arXiv (and prior to publication) \citet{poulin18} appeared on the arXiv. This paper presents a cosmological solution reconciling CMB, BAO, and Cepheid-calibrated supernova data. The solution is consistent with our analysis here: namely it has an early dark energy component contributing significantly in the scale factor window we have just described. It also leads to predictions that appear to be testable by future measurements of CMB polarization.

\section{Conclusions}

Following \citet{bernal16a} we have compared, using more recent data, empirical CDL determination of $r_{\rm s}$ with its inference assuming the \LCDM\ model and given a variety of CMB datasets. 
Casting the tension between the CDL and \LCDM\ + CMB datasets in terms of $r_{\rm s}$, as opposed to $H_0$, weakens the statistical significance, but helps to clarify the space of possible cosmologies that could reconcile these datasets. 
As the inverse distance ladder analyses have pointed out, modifying the shape of $D_A(z)$ at $z < 1$ can at most be a sub-dominant part of the solution. 

Because SNe cover the range of redshifts of the BOSS galaxy BAO data, our CDL inferences of $r_{\rm s}$ are highly model-independent. For the Spline model, which we prefer for this purpose over \LCDM\ due to its modest cosmological model assumptions\footnote{There is an implicit assumption of zero mean curvature. As discussed above, we expect that if we relaxed this assumption our results would only shift a small amount, as was the case for a similar analysis \citep{verde17}.}, from the Cepheid, SNe, and BAO datasets, we find $r_{\rm s} = 137.7 \pm 3.6$ Mpc. 

This result is 2.6$\sigma$ lower than the result from \planck{} TT+TE+EE+lowE (which we have referred to simply as {\it Planck}). We calculated the statistical significance of the difference between the CDL result and the \LCDM\ + CMB data results for a variety of CMB datasets and found they ranged from 2.1 to 3.0$\sigma$. Perhaps of particular interest, the combination of the highest precision non-\planck{} data, WMAP9 + SPT-SZ + ACT, gives an $r_{\rm s}$ that is 3.0$\sigma$ discrepant from the above CDL result. It is clear that the sound horizon differences cannot be explained by an unknown systematic error in the \planck\ data.

Expanding the model space to \LCDM\ + $N_{\rm eff}$ does not reduce the tension of the CDL $r_{\rm s}$ with the {\it Planck} $r_{\rm s}$. Although the error bar for the {\it Planck}-determined $r_{\rm s}$ increases considerably, the CDL error remains larger, and the central value for the {\it Planck}-determined $r_{\rm s}$ shifts to a slightly higher value. Expanding further to \LCDM\ + $N_{\rm eff} + Y_{\rm P}$ only reduces the tension from 2.6$\sigma$ to 2.3$\sigma$. The CMB data have no significant preference for these extensions.

While the CMB data show no preference for those particular extensions, we point out here that there are hints/weak evidence of inconsistencies of the CMB data with the \LCDM\ model. Parameter constraints derived from different angular scales, such as the \planck{} temperature power spectra at $\ell < 800$ compared to $\ell > 800$, are uncomfortably different, with a statistical significance that varies between 1.5 and 2.9$\sigma$ depending on details of the analysis and how the question of consistency is posed \citep{addison16, planck16-51, kable18}. Driven by small angular scales better measured by the South Pole Telescope, there is a 2.1$\sigma$ tension between SPT-SZ determination of cosmological parameters and those from \planck{} \citep{aylor17}. It is possible that these are hints relevant to the sound horizon discrepancy, but current data are not yet clear on the matter, and no model has been discovered, to our knowledge, that both addresses the sound horizon discrepancy and improves CMB internal consistency.

We argued that viable cosmological model solutions are likely to include important changes from \lcdm\ in the two decades of scale factor growth prior to recombination. This statement is interesting because it has an exciting implication: significant changes in this time period are likely to lead to consequences observable with near-future precision observations of CMB polarization. 

\comment{
Although we do not rule out the possibility of other types of solutions, the most direct solutions are those that differ significantly from the standard cosmological model in the two decades of scale factor expansion immediately prior to recombination. Changes that are only important earlier cannot reduce the sound horizon significantly since, in the standard cosmological model, near the best-fit location in parameter space given \planck\ data, greater than 95\% of $r_{\rm s}$ is generated in the final two decades of scale factor increase prior to recombination. Changes that are only important after recombination can not directly alter the sound horizon. We have not ruled out the possibility that post-recombination changes could adjust our inference of the matter density, and thereby adjust $r_{\rm s}$ indirectly, but our judgment is that such an approach is unlikely to succeed. Even if this judgment is incorrect, it appears likely to us that any viable cosmological solution will make predictions testable with improved measurements of CMB polarization -- either because the dynamics of observable scales have been altered prior to recombination, or because the post-recombination effects that have, to date, confused our inference of the matter density, are unlikely to mimic \lcdm\ well enough to escape detection under measurements of significantly higher precision. }

We produced forecasts for one such model adjustment: allowing for $N_{\rm eff}$ to be a free parameter, which directly alters pre-recombination dynamics. We found a three-fold improvement in the constraints on $r_{\rm s}$ when combining \planck{} with the SPT-3G \citep{benson14} dataset.
Whether or not the solution to the discrepancy is cosmological, we can expect future observations of the CMB from SPT-3G and other future CMB surveys like AdvACT \citep{henderson16}; Simons Observatory \citep{SO18}, CMB-S4 \citep{cmbs4-sb1}, and PICO \citep{young18}, to reveal further clues via their sensitivity to the acoustic dynamics of the plasma.

\appendix 
\section{Forecast Inputs}
\label{sec:forecastinputs}
In \S\ref{sec:results}, we presented the expected constraints on $r_{\rm s}$ that can be achieved by SPT-3G \citep{benson14} for two extensions of the \lcdm{} model: \lcdm+N$_{\rm eff}$ and \lcdm+N$_{\rm eff}$+$Y_p$. Here, we describe the inputs for the forecast.

SPT-3G is the third generation millimeter-wave camera on the South Pole Telescope~\citep{carlstrom02}. 
SPT-3G commenced operations in early 2018 and is currently observing a 1500 deg$^{2}$ sky-patch in the Southern hemisphere.
It is expected to achieve projected levels of noise in intensity maps of 3.0, 2.2, and 8.8 $\mu$K-arcmin at 95, 150, and 220 GHz respectively at the end of five years \citep{bender18}.
A primary goal of this SPT-3G survey is to produce a high signal-to-noise ratio CMB lensing map for delensing the BICEP Array \citep{hui18} observations that overlap with the SPT-3G 1500 deg$^2$ patch.
When completed, it will be the deepest high-resolution CMB survey of any patch of this size or larger.

For the Fisher forecast, we use TT, TE, EE, and $\phi\phi$ power spectra as inputs. 
We construct the covariance matrix assuming the T and E mode maps are fully delensed and therefore not correlated by lensing. 
To model the noise, we use the projected SPT-3G noise levels ($\sqrt{2}$ higher in polarization) to construct a foreground-reduced estimate of $N_{\ell}$ using the internal linear combination (ILC) method as described in \citet{raghunathan17}. 
We add a $1/\ell$ knee at $\ell_{\rm knee} = 1200, 2200, 2300$ for the three channels in T and $\ell_{\rm knee} = 300$ for channels in P to model large angular scale noise. 
For the lensing spectrum $C_{\ell}^{\phi\phi}$, we compute the noise with the minimum-variance combination of TT, TE, EE, TE, and EB quadratic estimators~\citep{hu02a}. 
We do not model the covariance of the common patch between \planck{} and SPT-3G because SPT-3G's patch is much smaller than {\it Planck}'s and the high S/N mode coverages for each experiments overlap little.

To include \planck\ constraints in the forecast we ``Fisher-ize" the \planck\ chain from the relevant parameter space: estimating the parameter covariance matrix from the chain and then inverting it to get the parameter Fisher matrix. We then add this to the SPT-3G Fisher matrix to get our final Fisher matrix. The \planck{} chains we use are for the data combination TT+TE+EE+lowE as explained in \citet{planck18-6}. 

\begin{acknowledgements}
We thank J. Bernal, L., Verde and D. Scolnic for useful conversations, and E. Calabrese for providing the ACTpol $r_{\rm s}$ constraint in Fig. 3. 
We use the CosmoSlik package\cite{2017ascl.soft01004M} to sample our results.

The work of KA and LK was supported in part by NSF support for the South Pole Telescope collaboration via grant OPP-1248097). The work of MJ was supported via the NSF REU program at UC Davis, via grant PHY-1560482. 
SR is supported by the Australian Research Council's Discovery Projects scheme via grant DP150103208.
WLKW was supported in part by the Kavli Institute for Cosmological Physics at the University of Chicago through an endowment from the Kavli Foundation and its founder Fred Kavli.
\end{acknowledgements}

\bibliography{spt.bib}

\end{document}